\input psfig.sty
\documentstyle{cupconf}

\ifoldfss
\else
  \ifnfssone
    \newmathalphabet{\mathit}
      \addtoversion{normal}{\mathit}{cmr}{m}{it}
      \addtoversion{bold}{\mathit}{cmr}{bx}{it}
    \newmathalphabet{\mathcal}
      \addtoversion{normal}{\mathcal}{cmsy}{m}{n}
    \else
    \ifnfsstwo
    \fi
  \fi
\fi

\def\hexnumber#1{\ifcase#1 0\or1\or2\or3\or4\or5\or6\or7\or8\or9\or
 A\or B\or C\or D\or E\or F\fi }

\makeatletter
\ifx\CUP@mtlplain@loaded\undefined
\else
\fi
\makeatother
 \makeatletter
 \ifx\CUP@mtlplain@loaded\undefined
   \font\tenbmi=cmmib10 at 10pt
   \font\sevenbmi=cmmib10 at 7pt
   \font\fivebmi=cmmib10 at 5pt

   \newfam\bmifam
   \textfont\bmifam=\tenbmi
   \scriptfont\bmifam=\sevenbmi
   \scriptscriptfont\bmifam=\fivebmi
   
 \fi
 \makeatother
\ifnfsstwo

\fi
\ifnfssone

\fi
\ifoldfss

\fi

\mathchardef\varLambda="0103

\makeatletter
\ifx\CUP@mtlplain@loaded\undefined
\else
\fi
\makeatother
\makeatletter
\ifx\CUP@mtlplain@loaded\undefined
  \font\tenbms=cmbsy10
  \font\sevenbms=cmbsy10 at 7pt
  \font\fivebms=cmbsy10 at 5pt
  \newfam\bmsfam
  \textfont\bmsfam=\tenbms
  \scriptfont\bmsfam=\sevenbms
  \scriptscriptfont\bmsfam=\fivebms

  \edef\bsy@{\hexnumber\bmsfam}
  \mathchardef\bnabla="0\bsy@72
\fi
\makeatother
\title[Are CSO host-galaxies undergoing merging?]{Are CSO host-galaxies undergoing merging?}

\author[Liu Xiang]%
{L\ls I\ls U\ns X\ls I\ls A\ls N\ls G\ls}

\affiliation{Urumqi Astronomical Observatory, NAOCAS, Urumqi, 830011, China} 

\begin{document}
\ifnfssone
\else
  \ifnfsstwo
  \else
    \ifoldfss
      \let\mathcal\cal
      \let\mathrm\rm
      \let\mathsf\sf
    \fi
  \fi
\fi

\maketitle

\begin{abstract}

The galaxy-type CSO (compact symmetric objects) rate from an intermediate large sample PR+CJ1 
is found to be consistent with the Sb-Sb merger rate of 4.6\% in redshift z$<$1.
 The CSO detection rate seems going down when the sample size is increasing, which could not reach
 the high merger rate in z$>$1, yet no deep VLBI sample in z$>$1 is available now.

\end{abstract}

\firstsection 
\section{Introduction}

The complete PR+CJ1 sample consists of 200 radio
sources with flux $>$0.7Jy at 5 GHz and declination $>$35 degrees.
 The 8/65 sources in PR and the 15/200 sources in PR+CJ1
sample are identified as CSOs (Polatidis et al. 1999), so the CSO rate is 12\% in
PR sample (flux $>$1.3Jy at 5 GHz), 7.5\% in PR+CJ1 sample.
The CSO rate from PR+CJ (including flat-spectrum samples CJ2 and CJF, 
with flux $>$0.35 Jy at 5 GHz) is 4.4\%(18/411) (Peck \& Taylor 2000).
 The COINS sample (CSOs observed in the northern sky, based on VLBI continuum surveys of nearly 2000
compact radio sources, i.e. PR+CJ+VCS, VCS:VLBA Calibrator surveys, with flux $>$100 mJy 
at 5 GHz, see Peck \& Taylor 2000)
gives a CSO rate of 2.1\%(39/1900). From the JVAS and CLASS flat-spectrum
 sample (with flux $>$100 mJy at 8.4 GHz) the CSO rate is 0.8\%(14/1665) (Augusto et al. 1999).
The CSO detection rate does appear going down when the sample size is increasing. 
However, both the uv-coverage and the sensitivity of the VCS are considerably worse than the 
PR and CJ surveys, it is possible that some CSOs have been missed in the VCS (Peck \& Taylor 2000).
 The flat-spectrum samples could also miss some CSOs, since most of CSOs from
PR+CJ1 sample show steep spectra (they are usually GPS sources).

\begin{table*}
\scriptsize
\begin{center}
\end{center}
\vspace{6pt}
\begin{tabular*}{\textwidth}{@{\hspace{\tabcolsep}
\extracolsep{\fill}}p{5pc}llllll}
\hline\hline\\[-6pt]
Name & z & ID & GPS & m$_{r}$ & Comments for the optical hosts\\
[4pt]\hline\\[-6pt]
0108+388 PR & 0.669 & G & Yes & 21.8 & 
very red color r-i=1.2 galaxy with an asymmetric\\
 &  &  &  &  &  morphology, the nucleus is more prominent in\\
 &  &  &  &  &  i band, suggesting a nuclear obscuration -[1]\\

0404+768 PR & 0.5985 & G &  Y &  20.7 & 
has weak diffuse structure, possible companion,\\ 
 &  &  &  &  & and very red color r-i=1.7 -[1] \\

0646+600 CJ1 & 0.455 & Q & Y-[2] & 19.9-[3] & 
a number of faint galaxies surrounded and a\\
 &  &  &  &  & possible companion -[3]\\

0710+439 PR & 0.518 & G & Y & 20.2 &    
very irregular morphology with very red (r-i=1.9)\\
 &  &  &  &  & color, there is evidence of strong interaction and\\
 &  &  &  &  & presence of large amount of obscuring matter;\\
 &  &  &  &  & many other resolved objects are present around\\
 &  &  &  &  & the parent object suggesting it is the dominant\\
 &  &  &  &  & galaxy in a cluster -[1]\\
1031+567 PR &  0.45 &  G & Y & 20.2  & elliptical-like galaxy showing a possible asymmetric\\
 &  &  &  &  & morphology, weak resolved objects are also present\\
 &  &  &  &  & in the field suggest it is the domination in a cluster -[1]\\

1225+368 CJ1 & 1.974 & Q & Y-[2] & 21.6-[2]\\

1242+410 CJ1 & 0.813 & Q & & 19.6-[4] & very red compact object -[5]\\

1311+678 CJ1 &      & EF &   & & steep spectrum with $\alpha_{(11-6)cm}=-0.73$ -[6]\\

1358+624 PR & 0.431 & G &  Y & 19.8 & a boxing morphology showing the 
obscuration\\
 &  &  &  &  & by dust, more visible in r band and a resolved\\
 &  &  &  &  & object is visible at 5" north, with r-i=0.3 -[1]\\

1437+624 CJ1 & 1.090 &  Q &   &   19.0-[7] &   \\

1843+356 CJ1 &  0.763 & G &  Y-[1] & 21.7 &       faint galaxy -[1]\\

1943+546 CJ1 & 0.263 &  G &  &  17.6-[8] &
apparently in a faint cluster, faint objects\\
 &  &  &  &  & surrounded -[8] and with $\alpha_{(11-6)cm}=-0.55$ -[6]\\

2021+614 PR & 0.2266 & G &  Y-[9] &  17.3 &
a low level extended halo around and the presence\\ 
 &  &  &  &  & of at least two candidate companions within\\
 &  &  &  &  & 12" of the galaxy -[2]\\

2342+821 PR & 0.735 & Q & Y-[1] & 20.1-[1] &
weak secondary emission with r-i=0.8 -[1]\\

2352+495 PR & 0.2383 & G & Y-[9] & 18.5 & 
an elliptical galaxy is located in a distant cluster, \\ 
 &  &  &  &  & nearby faint objects are candidate companions -[2]\\
\hline
\end{tabular*}
\vspace{6pt}

{\sl{Notes:} Source parameters if not referenced are from Polatidis et al. 1999
 and Snellen et al. 1996 (for m$_r$).
References: [1] Stanghellini et al. 1993, 1998.
[2] O'Dea et al. 1990. [3] Stickel \& K\"uhr 1993a. [4] Carballo et al. 1999.
[5] Vigotti et al. 1989. [6] Stickel \& K\"uhr 1994. [7] Aldcroft et al. 1994.
[8] Stickel \& K\"uhr 1993b. [9] Snellen et al. 1996}
\end{table*}

\section{Merger rate, CSO rate and discussion}
Gravitional interaction and mergers affect the the morphologies and dynamics
of galaxies from our Local Group to the limits of the observable universe.
Observations of interacting galaxies at low redshifts (z$<$0.2) yield detailed
information about many of the processes at work, these processes and the
growing evidence indicate that mergers play a major role in the delayed
formation of elliptical and early-type disk galaxies both in the field and
in clusters (Schweizer 1999). Merger ages from 65 E and S0 galaxies based on UBV
colors and a simple two-burst model of star formation spread over most
of the age of universe (Schweizer 1999). 
We measure the Sb-Sb merging (to E+S0, efficiency=0.10) rate from Schweizer \& Seitzer (1992, hereafter SS)
assuming that the epochs of 3, 6, 9, 12 and 15 Gyr ago are equal
to the time at redshift z=1,2,3,4,5 correspondingly. The result (i.e.the cumulative 
rate in per z/3 range) is (\%) 3.1, 0, 1.5, 4.6, 12.3, 9.2, 15.4, 26.2, 16.9, 7.7, 1.5, 1.5, 0
in redshift from 0 to 4.33 as shown in Fig.1. The Gaussion fitting formula:
y=1.93+20.63exp(-2.16(z-2.45)$^2$),
the fitting is not good for the data in range z=0-2, we can see that the data slope is
less steep than that of in z$>$2.4.
We then fit the data in z=0-2 with a power law (1+z)$^{2.28}$
. The evolution of the merger(or pair) rate in z$<$2 has been studied by
several authors (see Gottl\"ober et al. and references therein), which follows a 
power law (1+z)$^{3\pm1}$.
For comparing to the CSO rate (the CSO rate is indeed the cumulative rate in their 
corresponding z range as in Fig.1), we compute the cumulative merger rate in range z=0-1, z=1-2, 
z=2-3, z=3-4 and draw the dash line as in Fig.1.
We find that the galaxy CSO rate of 4.5\% in PR+CJ1 is close to the cumulative Sb-Sb merger rate
of 4.6\% in z$<$1. The CSO rate 7.5\% of PR+CJ1 (line 2 in Fig.1) in range z=0-2 
, we think, is not true for z=1-2, because only 2 CSOs (quasars with z=1.09 and 1.974) are detected in z=1-2. 
The CSO rate of 12\% in PR is much higher than the cumulative merger rate in z$<$1, this could be due to the
PR sample is still not a deep sample in z$<$1.
It suggests that in the intermediate deep sample PR+CJ1 (flux $>$0.7 Jy), the cumulative galaxy CSO rate
is consistent with the cumulative Sb-Sb merger rate in z$<$1. This should be investigated further in 
more deep and complete sample.
On the other hand, we collect the properties of the CSO hosts from literature as in the table above, the CSO 
hosts seem having the symptoms of merging or merging environment, however, this can not confirm 
if they are undergoing merging.

In the PR sample, 30/65 sources are identified as quasars, 29/65 are
galaxies, and 6/65 are BL Lac sources. The sources in PR+CJ1 are not completely identified.
From a large optical galaxy sample, 41\% of ellipticals, 65\% of galaxies and
 75\% of peculiars are detected in NVSS with $>$2.5 mJy
(Cotton and Condon 1998). The ellipticals and peculiars 
mostly are due to AGN activities, as well as the spiral galaxies
detected in NVSS are usually starburst ones.
At flux densities greater than 10 mJy, more than 95\% of the optical
identifications of radio sources are red ellipticals, Seyferts, and quasars, but
this fraction drops to 30\% at 5 mJy; the majority of sources with 200 $\mu$Jy$<$
 S$_{20cm}$$<$5 mJy have optically blue, disk-galaxy counterparts (Richards 1998).
So for CSO rate detection, VLBI samples can be as deep as $>$10 mJy without 
considering of starburst galaxies.

It is proposed that all elliptical galaxies are originated from spiral-spiral galaxy
merging (Schweizer 1997). 
We searched in NVSS for 32 ellipticals of SS
sample, 15/32(47\%) ellipticals show radio emission with $>$2.5 mJy.
More than 50\% ellipticals have not been detected in radio indicates that the radio phase of
 merger (which form elliptical eventually) is very short comparing with
the lifetime of elliptical. For example, the radio lobes of Cygnus A is aging at 
0.01 Gyr (Kaiser 2000), while the ages of young ellipticals in SS sample are greater than 0.9 Gyr.
Indeed, 2 merger remnants (0.3-0.6 Gyr) in SS sample are point-like radio sources in NVSS.
The CSO lifetime is less than 10$^{3-4}$years (Owsianik et al. 1999 and references therein; 
Xiang et al. 2000). CSOs may be triggered by  
the merging of nuclei of spirals, since strongly merging pairs we known have not shown 
jet-like radio emission, and their nuclei have not yet merged. So the CSOs could occur at
late merging stage (say at the
elliptical-forming stage, in that time, the nuclei of spirals is merging and the arms of spirals
are already merged).

 Quasar-type CSOs in PR+CJ1 show a large deviation from galaxy-type CSOs in Hubble diagram of Fig.2
. Radio selected QSOs are usually hosted by giant elliptical galaxies (Benn et al. 1998).
So they may be the mergers from giant spirals. We think their unusual luminosities and
 relatively less data points in Fig.2 result the flatter slope than that of galaxy CSOs.  
The scattering spectrum of QSOs in Fig.2 suggests the QSOs are not good standard candles, they may be
more bright virtually with increasing redshift.

It is expected that radio AGNs will live longer period in the MSO phase
 because MSOs are defined to have broad linear scales (1-15 kpc), we
should detect more MSOs in a complete sample e.g. PR+CJ1. However, the MSO detection
rate is 2\% in PR (one quasar) and 5\% in PR+CJ1 sample (Polatidis et al. 1999). 
One reason may be that
the symmetric two-side jets are much disturbed by the remnant of spiral arms, 
we assumed that the merging of spiral arms is finished when a CSO is 
triggered. At this point, most of CSS and some of GPS sources might be actually MSOs.\\
Liu Xiang acknowledge P. N. Wilkinson for illuminating discussions, thank Zhang Haiyan for help
on the figures.

\begin{figure}[!h]
\centerline{\psfig{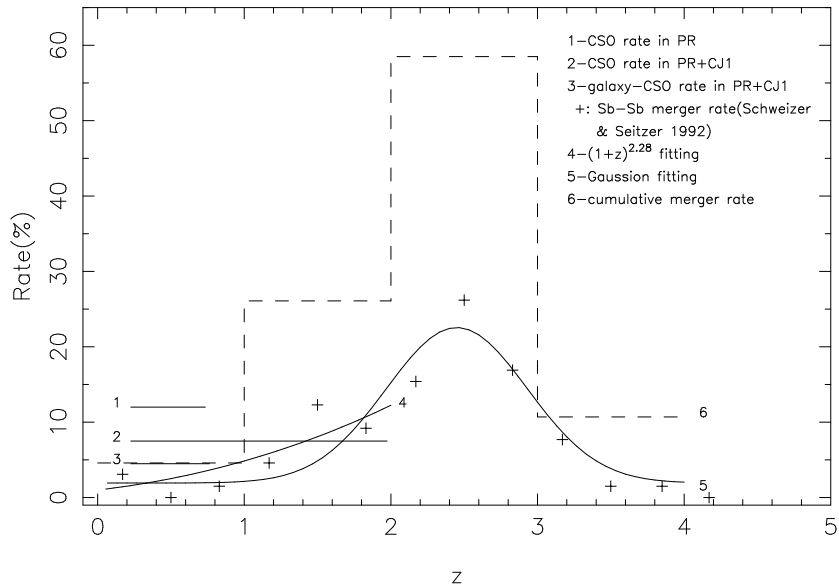}
}
\caption{The cumulative CSO rate and the Sb-Sb spiral merging (into E or S0) rate}
\end{figure}

\begin{figure}[!h]
\centerline{\psfig{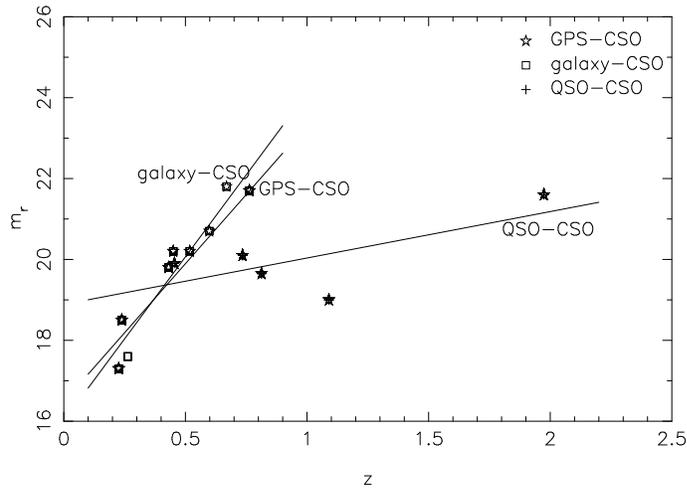}
}
\caption{Hubble diagram for CSOs in PR+CJ1 sample}
\end{figure}



{}
\end{document}